\documentclass[aps,prl,twocolumn,superscriptaddress,groupedaddress]{revtex4}  
\usepackage{graphicx}  
\usepackage{dcolumn}   
\usepackage{bm}        
\usepackage{amssymb,times}   

\begin{document}
\title{Characterization of multiple topological scales in multiplex networks through supra-Laplacian eigengaps}
\author{Emanuele Cozzo}
\affiliation{Institute for Biocomputation and Physics of Complex Systems (BIFI), University of Zaragoza, Zaragoza 50009, Spain}
\affiliation{Department of Theoretical Physics, University of Zaragoza, Zaragoza 50009, Spain}

\author{Yamir Moreno}
\affiliation{Institute for Biocomputation and Physics of Complex Systems (BIFI), University of Zaragoza, Zaragoza 50009, Spain}
\affiliation{Department of Theoretical Physics, University of Zaragoza, Zaragoza 50009, Spain}
\affiliation{Complex Networks and Systems Lagrange Lab, Institute for Scientific Interchange, Turin, Italy}

\date{\today}

\begin{abstract}
Multilayer networks have been the subject of intense research during the last few years, as they represent better the interdependent nature of many real world systems. Here, we address the question of describing the
three different structural phases in which a multiplex network might exist. We show that each phase can be characterized by the presence of gaps in the spectrum of the supra-Laplacian of the multiplex network. We
therefore unveil the existence of different topological scales in the system, whose relation characterizes each phase. Moreover, by capitalizing on the coarse-grained representation that is given in terms of quotient graphs, we explain the mechanisms that produce those gaps as well as their dynamical consequences.
\end{abstract}

\maketitle

\section{Introduction}
Multiplex networks are a particular setting of multilayer systems that can be represented using different layers, each one containing a network that accounts for a type of interaction. The layers are coupled together, since each node might participate in more than one layer or network  \cite{Kivela2014Multilayer}. As recently shown, many real world systems are better described in terms of these multiplex networks rather than in the traditional (single-layer) complex networks representation \cite{Boccaletti06}. This is, for instance, the case of on-line social networks (Facebook, Twitter, etc.), in which some users participate in more than one social network, or that of a biochemical system in which different signaling channels work in parallel, or that of a multimodal transportation system. Thus, multiplex networks represent systems in which there are several topological levels, that we call layers.

Single-layer complex networks \cite{Boccaletti06} exhibit non-traditional critical effects \cite{dorogovtsev2008critical} due to their extreme compactness together with their complex organization. A central theoretical question in the study of multilayer networks in general and multiplex networks in particular is whether critical phenomena will behave differently on them with respect to traditional ones. So far, theoretical studies have pointed out that this is indeed the case \cite{parshani2010interdependent,son2012percolation,Cozzo2013}. Moreover, as recently shown \cite{RadicchiArenas,radicchi2014driving}, multiplex networks might show different structural phases. Namely, under some conditions, the multiplex system might behave as one interconnected system, while in other conditions, the layers can become effectively disconnected and behave as if they were isolated \cite{gomez2013diffusion}. In this work, we show that there are three different topological scales that can be naturally identified in a multiplex network: they are associated to {\it i)} the individual layers; {\it ii)} the network of layers; and {\it iii)} the aggregate network. Additionally, we demonstrate that the connection between these scales in terms of the spectral properties of the parent multiplex network and its coarse-grained representations characterizes the aforementioned structural phases. 

\section{General definitions}
To start with, let us provide some definitions that will allow us to formally represent a multiplex network. As in the case of single-layer networks, we consider a set of nodes $V$ that represents the constituents of the system. In addition, in order to distinguish different types of interactions, we have to consider a set of layers $L=\{1,2,\dots,m\}$, in which each index $\alpha\in L$ represents a layer of interaction. Moreover, as a given node might or might not be present in a given layer, we define the ordered \textit{node-layer} pair $(u,\alpha)$ to indicate that node $u$ participates in layer $\alpha$. Formally, we consider de binary relation $G_P=(V,L,P)$, where $P\subseteq V\times L$, i.e., $P$ is the set of node-layer pairs existing in the system, which is in general a sub-set of all the possible node-layer pairs given $V$ and $L$. Thus, $(u,\alpha)\in P$ is the representative of node $u$ in layer $\alpha$. Furthermore, there is a special case of multiplex network that happens when each node $u\in V$ has a representative in each layer, i.e., when $P= V\times L$. We refer to this multiplex as a \textit{node-aligned multiplex} \cite{Kivela2014Multilayer}. Finally, we denote by $n=\mid V \mid$ the number of nodes, and by $N=\mid P\mid$ the number of node-layer pairs. 

Given the previous definitions, we next represent, for each layer, the connections between node-layer pairs by a graph in the same way a graph represents a single-layer network. The graph $G_\beta(V_\beta,E_\beta)$, where $V_\beta=\{(u,\alpha) \in P \mid \alpha=\beta\}$, represents the interactions in layer $\beta$ between the node-layer pairs that are the representatives of the nodes in that layer. In other words, there is a link between $(u,\beta)$ and $(v,\beta)$ in $G_\beta$ if and only if node $u$ and node $v$ have an interaction of the type $\beta$. Each graph of this type will be called a \textit{layer-graph}. Thus, it follows that we can define the union of all layer-graphs, i.e., $G_l=\bigcup_\alpha G_\alpha$ and we call it the intra-layer graph. The intra-layer graph $G_l$ represents all the interactions between the representatives of all nodes in all layers.
In addition, we have to consider the couplings between node-layer pairs that represent the same node in different layers. To this end, we introduce the coupling graph $G_C$ on $P$ in which there is an edge between two node-layer pairs $(u,\alpha)$ and $(v,\beta)$ if and only if $u=v$, that is, when the two node-layer pairs represent the same node but in different layers. It is easy to realize that the coupling graph $G_C$ is always a union of cliques and of isolated nodes, in particular each clique is formed by all the node-layer pairs representing the same node. In the case of node-aligned multiplex networks, since each node has a representative in each layer, the cliques are all of the same size. Finally, a synthetic representation of the whole multiplex network can be defined as $G_\mathcal{M}=G_l\cup G_C$, i.e., the union of the intra-layer graph and the coupling graph. This graph is called a \textit{supra-graph}. A supra-graph hence represents a multiplex network in the same way that a graph represents a traditional single-layer network. As usual, an adjacency matrix or a Laplacian matrix can be associated to the supra-graph $G_\mathcal{M}$. In this paper we focus on the Laplacian and refer to it as the supra-Laplacian \cite{gomez2013diffusion}. 

If $\mathbf{L}^\alpha$ is the Laplacian associated to a layer-graph $G_\alpha$ and $\mathcal{L}_C$ is the Laplacian associated to the coupling graph $G_C$, the supra-Laplacian associated to $G_\mathcal{M}$ can be written as
\begin{equation}
\bar{\mathcal{L}}=\bigoplus_\alpha \mathbf{L}^\alpha +\mathcal{L}_C,
\end{equation}

\begin{figure}[t]
\centering
\includegraphics[width=\columnwidth,angle=0]{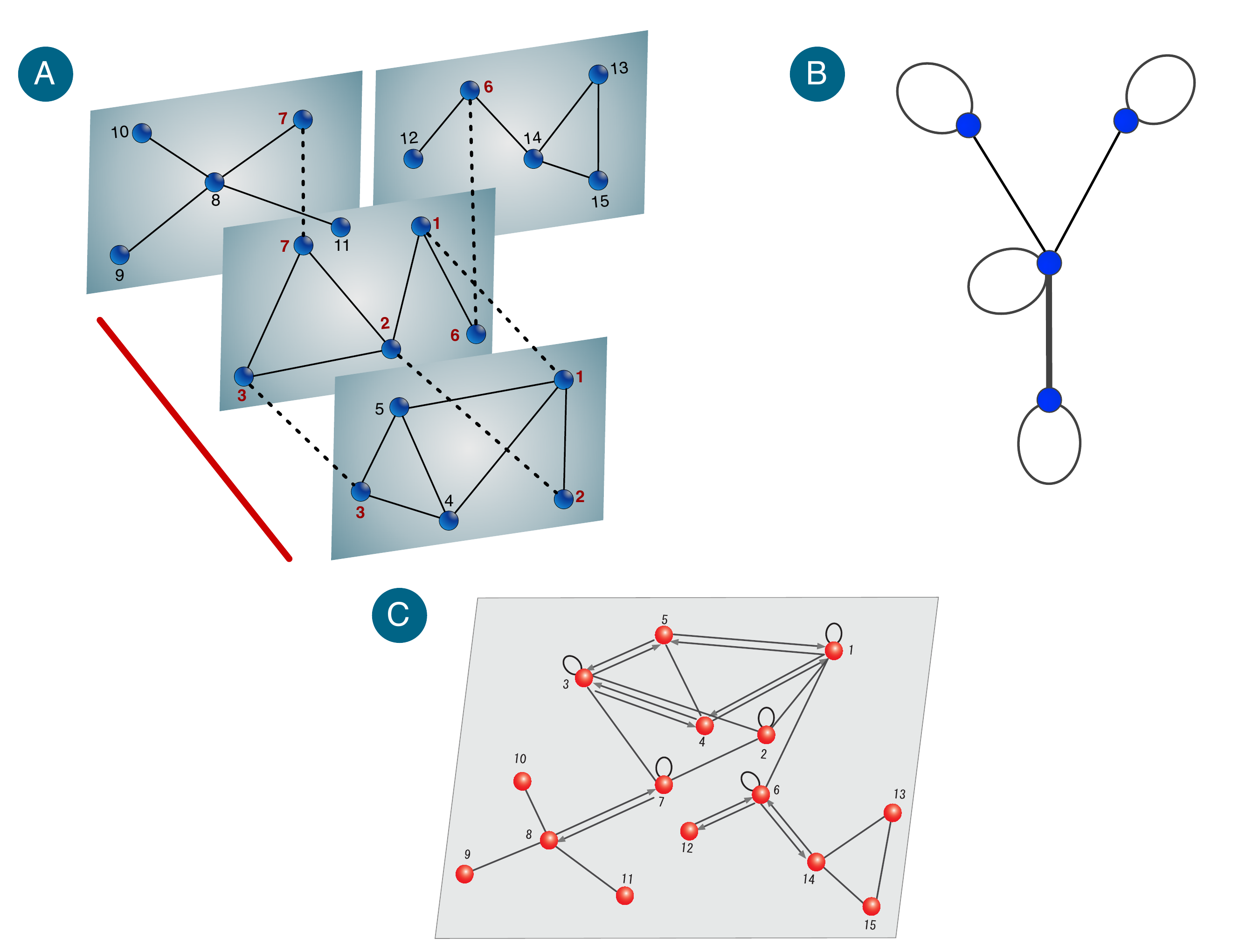}
\caption{(color online) Schematic representation of a multiplex network. Panel (a) shows an example of a multiplex network. The two other panels depict its corresponding coarse-grained reductions: the network of layers (b), and the aggregate network (c). Details of how these reductions are obtained can be found in \cite{sanchez2014dimensionality}.}
\label{fig1}
\end{figure}

In order to make this work self-contained, we end this section introducing two coarse-grained reductions of a multiplex network that were previously defined in \cite{sanchez2014dimensionality}, namely the \emph{aggregate network} and the \emph{network of layers}, see Fig.\ \ref{fig1}. Both are based on the notion of quotient graphs, resulting in an exact relation between their adjacency and Laplacian spectra and those of the parent multiplex network. Roughly speaking, in the aggregate network a link exists from node $u$ to node $v$ if and only if they are connected in at least one layer and the link is weighted by the number of links they have over the number of layers in which $u$ is a representative. As one can easily realize, the aggregate network is in general directed as far as the multiplex network is not node-aligned. In the network of layers, there is a node for each layer and a link from layer $\alpha$ to layer $\beta$ is weighted by the number of nodes they share over the number of node-layer pairs in $\alpha$. In the case of node-aligned multiplex networks, the network of layers is a complete graph and all links have weight equal to one. In both coarse grained networks, nodes have self-loops weighted with the same rules.
Let $\tilde{\mathbf{L}}_a$ and $\tilde{\mathbf{L}}_l$ be the Laplacian of, respectively, the aggregate network and of the network of layers. They are given by
\begin{eqnarray}
\tilde{\mathbf{L}}_a=\Lambda_n^{-1}\mathbf{S}_n^T\bar{\mathcal{L}}\mathbf{S}_n\\
\tilde{\mathbf{L}}_l=\Lambda_l^{-1}\mathbf{S}_l^T\bar{\mathcal{L}}\mathbf{S}_l,
\end{eqnarray}
where $\Lambda_n=diag(\kappa_1,\dots,\kappa_n)$ is the multiplexity degree matrix, i.e., it has the number of layers in which a node has a representative on the diagonal, $\mathbf{S}_n=(s_{iu})$ is the node characteristic matrix whose elements $s_{iu}=1$ if and only if the node-layer pair $i$ is a representative of node $u$, $\Lambda_l=diag(n_1,\dots ,n_m)$ is the layer size matrix, i.e., it has the number of node-layer pairs of each layer on the diagonal, and $\mathbf{S}_l=(s_{i\alpha})$ is the layer characteristic matrix whose elements $s_{i\alpha}=1$ only if the node-layer pair $i$ is in layer $\alpha$.

\section{Characterization of multiple topological scales in multiplex networks}

In this paper, we focus our attention on the spectra of the supra-Laplacian and show how the interplay between different topological scales affects the whole structural organization of a multiplex network. The spectrum of the Laplacian is a natural choice to address this problem, since it reveals a number of structural properties. In particular, gaps in the Laplacian spectrum (eigengaps) are known to unveil a number of structural and dynamical properties of the network related to the presence of different topological scales in it, from communities at different topological scales to synchronization patterns \cite{shen2010covariance,review_sync}. Thus, the emerging of an eigengap points to structural changes that result in qualitatively different dynamical patterns. For this reason, we introduce a weight parameter $p$ that allows us to tune the relative strength of the coupling with respect to the intra-layer connectivity. The parameter $p$ appears naturally as a physical parameter when one considers, for instance, a diffusion dynamics \cite{gomez2013diffusion} or a spreading process \cite{Cozzo2013}, thus connecting topology and dynamics.

The supra-Laplacian with the weight parameter $p$ reads as:
\begin{equation}
\bar{\mathcal{L}}=\bigoplus_\alpha \mathbf{L}^\alpha +p \mathcal{L}_C,
\end{equation}
where $\mathbf{L}^\alpha$ is Laplacian of the layer-graph $G_\alpha$, while $\mathcal{L}_C$ is the Laplacian of the coupling graph. 

Denote the eigenvalues of the Laplacian of the aggregate network, $\tilde{\mathbf{L}}_a$, as $0=\tilde{\mu}^{(a)}_{1}<\tilde{\mu}^{(a)}_{2}\leq \dots\leq\tilde{\mu}^{(a)}_{n}$ and that of the network of layers, $\tilde{\mathbf{L}}_l$, as $0=\tilde{\mu}^{(l)}_{1}<\tilde{\mu}^{(l)}_{2}\leq \dots\leq\tilde{\mu}^{(l)}_{m}$. From \cite{sanchez2014dimensionality}, we have that the eigenvalues of the coarse-grained networks interlace those of the parent supra-Laplacian, that is, 
\begin{eqnarray}
\bar{\mu}_i\leq\ \tilde{\mu}^{(a)}_{i}\leq \bar{\mu}_{i+N-n}\nonumber\\
\bar{\mu}_i\leq\ \tilde{\mu}^{(l)}_{i}\leq \bar{\mu}_{i+N-m}.
\end{eqnarray}
In particular, when the multiplex network is node-aligned, the spectrum of the Laplacian of the network of layers is a subset of the spectrum of the parent supra-Laplacian.

\begin{figure}[t]
\centering
\includegraphics[width=\columnwidth]{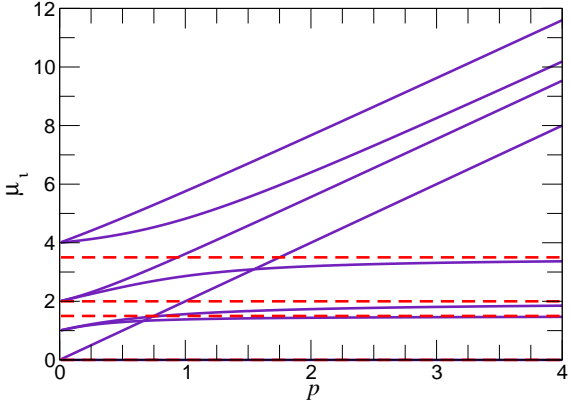}
\caption{(color online) Eigenvalues of a toy 2-layer multiplex with 4 nodes per layer as a function of $p$. Continuous blue lines are the eigenvalues of the multiplex network, whereas the dashed red lines are the eigenvalues of the aggregate network.}
\label{toy}
\end{figure}

Figure \ref{toy} shows the full spectrum of a toy node-aligned multiplex network of $4$ nodes and $2$ layers -thus $8$ node-layer pairs. We first note, as claimed in \cite{gomez2013diffusion,sole2013spectral}, that the spectrum splits into two groups: one made up by eigenvalues that remain bounded while increasing $p$, and another group of eigenvalues that diverge linearly with $p$. The whole characterization of the structural changes in a multiplex network basically depends on this splitting, i.e., on the emerging of gaps in the spectrum. 
The Laplacian spectrum of the network of layers is composed of just two eigenvalues: $0$ with multiplicity $1$, and $mp$ with multiplicity $(m-1)$. Because of the inclusion relation \cite{sanchez2014dimensionality} between the coarse-grained and the parent spectra, $mp$ will be always an eigenvalue of the supra-Laplacian. It results that, for low enough values of $p$, $mp$ will be the smallest non-zero eigenvalue of $\bar{\mathcal{L}}$. On the other hand, each eigenvalue $\bar{\mu}_i$ of $\bar{\mathcal{L}}$, with $i=1\dots n$, will be bounded by the respective Laplacian eigenvalue $\tilde{\mu}^{(a)}_{i}$ of the aggregate network because of the interlace. It is evident that, by increasing $p$, at some value $p=p^*$, it will happen that $\bar{\mu}_2\neq mp$ and that it will approach its bound $\tilde{\mu}^{(a)}_{2}$. For continuity, at $p^*$, $\bar{\mu}_3=mp$ must hold, since $mp$ is always an eigenvalue of the supra-Laplacian. $p=p^*$ is the point at which the structural transition described in \cite{RadicchiArenas,martin2014algebraic}  occurs, as already noted by Darabi Sahneh et al. \cite{sahneh2014exact}. 
Each eigenvalue up to $\bar{\mu}_n$ will follow the same pattern, following the line $\bar{\mu}_i=mp$ and departing from it to approach its bound $\tilde{\mu}^{(a)}_i$ when it hits the next eigenvalue $\bar{\mu}_{i+1}=mp$ (see Fig.\ \ref{toy}). The last eigenvalue crossing will be at the point $p=p^\diamond$ at which $\bar{\mu}_n=mp$, after that point $ \bar{\mu}_{n+1}=mp$ must hold and for continuity it will hold forever, since $\bar{\mu}_{n+1}$ is not bounded. To summarize, \textit{two structural transition points} are defined: one at the first eigenvalue crossing $mp=\bar{\mu}_3$ and another one at the last eigenvalue crossing $mp=\bar{\mu}_n$.

Continuing with this reasoning, it follows that the supra-Laplacian spectrum for $p>p^\diamond$ can be divided into two groups: one of $n$ bounded eigenvalues that will approach the aggregated Laplacian eigenvalues as $p$ increases,  and one of $N-n=n(m-1)$ eigenvalues diverging with $p$. Therefore, the system can be characterized by an eigengap emerging at $p^\diamond$. Moreover, while the splitting of the eigenvalues in these two groups is always present (because of the interlacing), the crossing of the eigenvalues at $p^*$ and at $p^\diamond$ (and between those points) only happens when the multiplex is node-aligned, this is because the inclusion relation only holds in that case.

In order to quantify an eigengap, we introduce the following metric:
\begin{equation}
g_{k}=\frac{\log(\bar{\mu}_{k+1})-\log(\bar{\mu}_{k})}{\log(\bar{\mu}_{k+1})}
\label{eigengap}
\end{equation}
and we will focus on $g_{n}(p)$, i.e., the gap emerging between the last bounded eigenvalue and the first unbounded at $p^\diamond$. By construction 
\begin{equation}
g_{n}(p^\diamond) = 0. 
\end{equation}
For $p>p^\diamond$, $\log(\bar{\mu}_{n+1})$ will diverge while $\log(\bar{\mu}_{n})$ will remain bounded by $\tilde{\mu}^{(a)}_{n}$, so $g_{n}$ will approach $1$. For $p<p^\diamond$, in general, both $\bar{\mu}_{n+1}$ and $\bar{\mu}_{n}$ will be in the continuous part of the spectrum in the large size limit, so $g_n$ will be 0 in this limit. That is, in the large system size limit,  
\begin{eqnarray}
g_n=0, p\leq p^\diamond\nonumber\\
g_n\neq 0, p>p^\diamond .
\end{eqnarray}
\begin{figure}
\includegraphics[width=8cm]{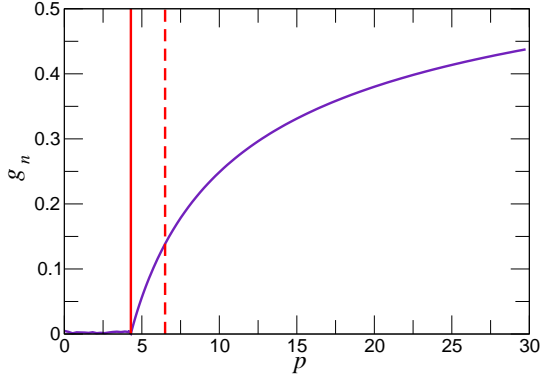}
\caption{(color online) Variation of the eigengap metric $g_n$ [Eq.\ (\ref{eigengap})] with $p$. The figure represents the eigengap between the last bounded and the first unbounded eigenvalue for a node-aligned multiplex network made up by two Erdos-Renyi networks of 200 nodes and an average degree $\langle k \rangle=5$. The vertical continuous line is the analytical value of $p^\diamond$, while the dashed line is the bound provided in the SM, Eq.\ (17).} 
\label{g_n}
\end{figure}
This phenomenology is confirmed by numerical simulations, see Fig. \ref{g_n}. It describes a structural transition occurring at $p^\diamond$. In the case of a non node-aligned multiplex network, where $p^\diamond$ is not defined since there is no crossing, $g_n(p)$ can be used to define it operationally. The exact value of $p^\diamond$ can be derived following \cite{sahneh2014exact} to be
\begin{equation}
p^\diamond=\frac{1}{2}\lambda_n(\mathbf{Q})
\label{exact}
\end{equation}
being, for the case of two layers, $\mathbf{Q}=\mathbf{L}^+ - \mathbf{L}^- {\mathbf{L}^+}^\dagger L^-$, $L^\pm=\frac{1}{2}(L_1\pm L_2)$, and $A^\dagger$ the Moore-Penrose pseudoinverse of $A$.

Interesting enough, an upper bound for $p^\diamond$ can be given in terms of the structural properties of the layers. In fact, a trivial bound, by definition, is given by
\begin{equation}
p^\diamond\leq \frac{\tilde{\mu}^{(a)}_n}{m}.
\label{pdiamondbound}
\end{equation}
The relation (\ref{pdiamondbound}) just states that $p^\diamond$ is defined as the value of $p$ at which the eigenvalue $mp$ exceeds the last bounded eigenvalues, the value of that bound being given by the largest Laplacian eigenvalue of the aggregate network. We can bound $\tilde{\mu}^{(a)}_n$ in terms of the structural properties of the layers. If $\omega^{(\alpha)}_i$ is the degree of node $u$ in layer $\alpha$, its strength in the aggregate network is $\tilde{\omega}_i=\frac{1}{m}\sum_\alpha\omega^{(\alpha)}_i$. Next let's define $\tilde{\omega}_{ij}=\tilde{\omega}_{i}+\tilde{\omega}_{j}$, $\forall i\sim j$, where $i\sim j$ indicates a link between $i$ and $j$ in the aggregate network. We have that \cite{kinkar2005}
\begin{equation}
\tilde{\mu}^{(a)}_n\leq \text{max}_{i\sim j}\{\tilde{\omega}_{ij} \},
\end{equation} 
resulting in the following bound for $p^\diamond$
\begin{equation}
p^\diamond \leq \frac{\text{max}_{i\sim j}\{\tilde{\omega}_{ij}\}}{m}=\frac{\text{max}_{i\sim j}\{\sum_\alpha{\omega}^{\alpha}_{ij}\}}{m^2}
\label{pdiamondbound2}
\end{equation} 

\section{The case of identical layers}
It is instructive to consider the special case of a multiplex network made up of layers that are all identical. Let $\mathbf{L}$ be the Laplacian matrix of all the layer-graphs. By definition, the supra-Laplacian can be written as
\begin{equation}
\bar{\mathcal{L}}=\mathbf{I}_m \otimes \mathbf{L} + p\mathbf{L}(\mathbf{K}_m)\otimes\mathbf{I}_n,
\label{identicalsupraL}
\end{equation}
where $\mathbf{L}(\mathbf{K}_m)$ is the Laplacian of the complete graph on $m$ nodes. 

Formally speaking, the multiplex network composed of layers that are all identical is given by the Cartesian product between the layer-graph and the network of layers. By definition, its Laplacian spectrum is given by all the possible sum between the eigenvalues of the Laplacian of the layer-graph and the eigenvalues of the networks of layers, i.e.,
\begin{equation}
\sigma(\bar{\mathcal{L}})=\{\mu_i(\mathbf{L})+\mu_k(\mathbf{L}(\mathbf{K}_m)) \mid i=1,\dots,n\ k=1,\dots,m\}.
\end{equation}
At $p=0$, all the eigenvalues are degenerated and the spectrum is composed by the eigenvalues of $\mathbf{L}$ with a multiplicity equal to the number of layers. For $p>0$, we have a set of $n$ constant eigenvalues that are equal to the eigenvalues of $\mathbf{L}$ and a set of $N-n$ eigenvalues of the form $\bar{\mu}_i=\mu_i(\mathbf{L}) + mp$. The two transition points can be calculated in this case, since it is easy to see that they are the points at which $mp$ intersects $\mu_2(\mathbf{L})$ and $\mu_n(\mathbf{L})$ respectively, i.e.,

\begin{equation}
p^*=\frac{\mu_2(\mathbf{L})}{m}
\end{equation}
and
\begin{equation}
p^\diamond=\frac{\mu_n(\mathbf{L})}{m}.
\label{identicaldiamond}
\end{equation}

\section{The Aggregate-Equivalent Multiplex Network}
To further characterize this transition, we would like to compare a multiplex network $\mathcal{M}$ with the coarse-grained networks associated to it. However, a direct comparison is not possible, since those structures have different dimensionalities. To overcome this problem, inspired by the case of identical layers, we define an auxiliary structure whose structural properties are completely defined by the aggregate network and the network of layers, but that has the same dimensionality of $\mathcal{M}$. We call it the Aggregate-Equivalent Multiplex (AEM). The AEM of a parent multiplex network $\mathcal{M}$ is a multiplex network with the same number of layers of $\mathcal{M}$, each layer being identical to the aggregate network of $\mathcal{M}$. Additionally, node-layer pairs representing the same nodes are connected with a connection pattern identical to the network of layers. Formally speaking, the AEM is given by the Cartesian product between the aggregate network and the network of layers. Thus, its adjacency matrix is given by
\begin{equation}
\mathcal{\mathbf{A}}=\mathbf{I}_m \otimes  \tilde{\mathbf{A}} + p\mathbf{K}_m\otimes \mathbf{I}_n, 
\end{equation}
where $\tilde{\mathbf{A}}$ is the adjacency matrix of the aggregate network, and its Laplacian matrix is given by
\begin{equation}
\mathcal{\mathbf{L}}=\mathbf{I}_m \otimes  \tilde{\mathbf{L}}_a + p\tilde{\mathbf{L}}_l\otimes \mathbf{I}_n,
\end{equation}   
where $\tilde{\mathbf{L}}_a$ is the Laplacian matrix of the network of layers and $\tilde{\mathbf{L}}$ is the Laplacian of a complete graph of $m$ nodes. Its Laplacian spectrum is fully determined in terms of the spectra of $\tilde{\mathbf{L}}_a$ and of the spectra of $\tilde{\mathbf{L}}_l$. In particular, we have
\begin{equation}
\sigma(\mathcal{\mathbf{L}})=\{\tilde{\mu}_a + \tilde{\mu}_l \mid \tilde{\mu}_a \in \sigma(\tilde{L}_a), \tilde{\mu}_l \in \sigma(\tilde{L}_l) \}.
\end{equation}
That is, each eigenvalue of $\mathcal{\mathbf{L}}$ is the sum of an eigenvalue of $\tilde{\mathbf{L}}_a$ and an eigenvalue of $\tilde{\mathbf{L}}_l$. We also note that, since $0$ is an eigenvalue of both coarse-grained Laplacians, the spectrum of both $\tilde{\mathbf{L}}_a$ and $\tilde{\mathbf{L}}_l$ are included in the spectrum of ${\mathbf{L}}$.

To compare the parent multiplex network with its AEM, we compute the quantum relative entropy between the former and the latter. The quantum entropy (or Von-Neumann entropy) of $\mathcal{M}$ is defined as
\begin{equation}
S_q(\mathcal{M})=Tr (\rho\log\rho) 
\end{equation} 
where $\rho=\frac{\bar{\mathbf{L}}}{2E+N(m-1)p}$, with $E$ being the number of intra-layer links in $\mathcal{M}$ \cite{passerini2009quantifying}, i.e., $\rho$ is the supra-Laplacian normalized by the degree sum. Thus, the quantum relative entropy of the multiplex network $\mathcal{M}$ with its associated AEM is defined as
\begin{equation}
R_q(\mathcal{M}\mid\mid AEM(\mathcal{M}))=Tr\rho(\log\rho-\log\sigma),
\end{equation}
where $\sigma$ is the supra-Laplacian of the AEM normalized by its degree sum without considering self-loops. It is worth noticing that the quantum relative entropy between a multiplex network of identical layers and its AEM is $0$ whatever the value of the coupling $p$ is (see the SI).
\begin{figure}[t]
\includegraphics[width=\columnwidth]{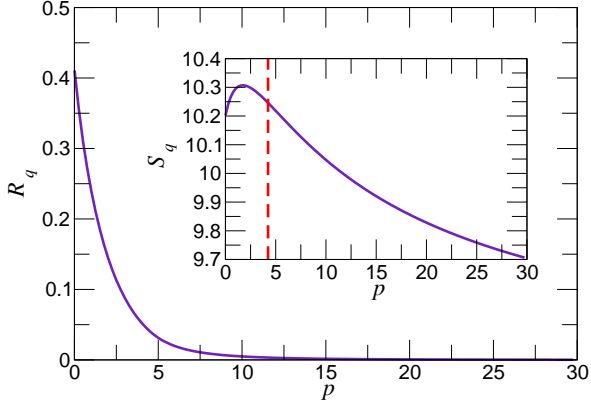}
\caption{(color online) Variation of the entropic measures with $p$. The figure shows the behavior of the relative entropy ($\times 10$), main panel, and of the relative quantum entropy (inset) when $p$ is increased. The multiplex network is the same in Fig.\ \ref{g_n} and the vertical line indicates the exact transition point $p^\diamond$.}
\label{entr}
\end{figure}

In Fig.\ \ref{entr} we show the quantum relative entropy between the parent multiplex and its AEM: it goes to $0$ when $p$ increases, which means that the parent multiplex will be indistinguishable from the AEM. Finally, it is informative to look at the quantum entropy of $\mathcal{M}$. $S_q(\mathcal{M})$ shows a clear peak after $p^*$ and before $p^\diamond$ (see the inset of Fig.\ \ref{entr}), i.e., in the region after the transition observed in \cite{RadicchiArenas,martin2014algebraic} and before the one we have introduced here. In fact, by studying the sign of the derivative of $S_q$, it can be proven that the quantum entropy must have a peak before $p^\diamond$.

\section{Conclusions}
To gain intuition on the phenomenological implications of our findings, it is enlightening to consider a diffusion dynamics. First of all, considering a diffusion dynamics, we can give a physical meaning to the coupling parameter $p$, i.e., assume that the diffusion constant for intra-layer diffusion is $D_{intra}$ while the diffusion constant for inter-layer diffusion is $D_{inter}$, then $p=\frac{D_{inter}}{D_{intra}}$ and the diffusion equation, after a rescaling of time, reads:
\[
\dot{\mathbf{x}}=-\bar{L}\mathbf{x}=-\bigoplus_\alpha\mathbf{L}^\alpha\mathbf{x}-p\mathcal{L}_C\mathbf{x}.
\] 
as in \cite{gomez2013diffusion}. In general, the physical meaning of the parameter $p$ depends on the actual system under study, however, it always represents the relative strength of the coupling between different node-layer pairs representing the same node in different layers with respect to the strength of the coupling of a node-layer pair with its neighbors in a given layer.

In diffusion dynamics, the large time scale is dominated by the bounded group of eigenvalues for $p\geq p^\diamond$. These eigenvalues are close to those of the aggregate network, indicating that each layer shows practically the same behavior of the latter network. This is because the fast time scale is dominated by the diverging group of eigenvalues that are close to those of the aggregate network plus those of the network of layers. In summary, the network of layers determines how each node-layer pair accommodates with its replica on a fast time scale, being always ``at equilibrium'', while the aggregate network determines how and on what time scale the global equilibrium is attained. From this point of view, the ``world'' will look the same from each layer and it will look like in the aggregate network. From the viewpoint of a random walk, we can look at the average commute time $c(i,j)$, i.e., the mean time needed by a walker starting in $i$ to hit node $j$ for the first time and come back. This quantity can be expressed in terms of the eigenvalue of $\bar{L}^\dagger$, the pseudoinverse of the supra-laplacian. Since the eigenvalues of $\bar{L}^\dagger$ are the reciprocal of the eigenvalues of $\bar{L}$, the aggregate network mean commute time $\tilde{c}(i,j)$ is a good approximation of $c(i,j)$ after $p^\diamond$\cite{saerens2004principal}:
\begin{equation}
\parallel c(i,j)-\tilde{c}(i,j)\parallel \leq E \frac{n(m-1)}{2p}.
\end{equation}

In summary, in this paper, capitalizing on the coarse-grained representations of a multiplex network via the aggregate network and the network of layers introduced in \cite{sanchez2014dimensionality}, we have unveiled the following structural phases as a function of $p$: before $p^*$ the system is structurally dominated by the network of layers, whereas after $p^\diamond$ it is structurally dominated by the aggregate network. Between these two points the system is in an effective multiplex state, i.e., neither of the coarse-grained structures dominate. In this region the VN-entropy -a measure of structural complexity - shows a peak. We have also shown that the novel structural transition at $p^\diamond$ is rooted in a gap that appears between the $n$-th and the $(n+1)$-th$=mp$ eigenvalues of the supra-Laplacian, while the transition at $p^*$ is rooted in a gap that disappears between the $2$nd$=mp$ and the $3$rd  eigenvalues of the supra-Laplacian. Finally, the definition of the Aggregate-Equivalent Multiplex allowed to compare the multiplex with its associated coarse-grained representations and to show that the relative entropy between the parent multiplex and its AEM varies smoothly with $p$, which implies that the two transitions are smooth from a global point of view. Altogether, the present work provides a full understanding of the spectrum of the supra-Laplacian of a multiplex network.

\begin{acknowledgments}
E. C was supported by the FPI program of the Government of Arag\'on, Spain. Y. M. acknowledges support from the Government of Arag\'on, Spain through a grant to the group FENOL, by MINECO and FEDER funds (grant FIS2014-55867-P) and by the European Commission FET-Proactive Project Multiplex (grant 317532).
\end{acknowledgments}

\end{document}